\documentclass{article}
\usepackage{graphicx}
\usepackage{url}

\newenvironment{pf}{\unskip{\bf Proof:}}{\unskip{\hfill $\Box$}}

\newcommand{\lemlab}[1]{\label{lemma:#1}}

\newcommand{\figlab}[1]{\label{fig:#1}}

\newcommand{\figref}[1]{\ref{fig:#1}}
\newcommand{\eqref}[1]{(\ref{eq:#1})}

\newtheorem{theorem}{Theorem}[section]
\newtheorem{lemma}[theorem]{Lemma}

{\catcode`\@=11
\gdef\setft#1#2#3{%
\def\@oddfoot{
{\setbox0=\hbox{#1}
\setbox1=\hbox{#3}
\ifdim\wd0>\wd1
\dimen0=\wd0
\box0\hfil#2\hfil\hbox to\dimen0{\hfil\hfil\box1}
\else \dimen0=\wd1
\hbox to\dimen0{\box0\hfil }\hfil#2\hfil\box1 \fi
}}} }

\def\complaint#1{}
\def\withcomplaints{
\newcounter{mycomplaints}
\def\complaint##1{\refstepcounter{mycomplaints}%
\ifhmode%
\unskip%
{\dimen1=\baselineskip \divide\dimen1 by 2 %
\raise\dimen1\llap{\tiny -\themycomplaints-}}\fi%
\marginpar{\tiny [\themycomplaints]: ##1}}%
}

\usepackage{amssymb}
\usepackage{amsmath}

\title{Computational Geometry Column 40}
\author{%
Joseph O'Rourke\thanks{
Dept. of Computer Science, Smith Col\-lege, North\-ampton, 
MA 01063, USA.
\-orourke@cs.\-smith\-.edu.
Supported by NSF Grant CCR-9731804.
}
}
\date{}
\begin{document}
\maketitle
\pagestyle{empty}
\thispagestyle{empty}

\begin{abstract}
It has recently been established 
by Below, De Loera, and Richter-Gebert 
that finding
a minimum size (or even just a small) triangulation
of a convex polyhedron is NP-complete.
Their 3SAT-reduction proof is discussed.
\end{abstract}

All triangulations of a polygon of $n$ vertices use the same number
of triangles, $n-2$, but the same does not hold in higher dimensions,
even for convex polytopes.
We will only discuss $3$-polytopes, i.e., convex polyhedra $P$,
where 
a {\em triangulation\/}
is a collection of
tetrahedra whose vertices are drawn from the vertices of $P$,
whose union is $P$,
and such that the intersection of any two of the tetrahedra
is either empty or a vertex, edge, or face common to the 
two tetrahedra.
The number of tetrahedra in 
a triangulation of a convex polyhedron of $n$ vertices might
be as small as $n-3$ or as large as $\binom{n}{2} -2n + 3$.
It is easy to obtain a linear-size triangulation
of a {\em simplicial\/}
polyhedron (all faces triangles)
by the following {\em starring\/} procedure (\cite[p.~424]{b-t-97}).
Select one vertex $v$ and include the tetrahedron
formed by
the convex hull of $v$ and each face $f$ not incident to $v$.
Because a simplicial 
polyhedron
has $F=2n-4$
faces, this method yields at most $2n-7$ tetrahedra
(at least three tetrahedra are incident to $v$).
For nonsimplicial polyhedra, it could be better.
For example, applied to a cube, starring results in a triangulation
by $6$ tetrahedra
(Fig.~\figref{cubes}a).
So this method provides a triangulation using at most twice
the minimum number.
But getting closer to the optimum (in the case of a cube, $5$
(e.g., Fig.~\figref{cubes}b)) has proven difficult.
Now we know why:
Below, De Loera, and Richter-Gebert (BDR) proved
that deciding whether a convex polyhedron can be triangulated
with fewer than $k$ tetrahedra is 
NP-complete~\cite{blr-cfst-00}.
\begin{figure}[htbp]
\centering
\includegraphics[width=0.60\linewidth]{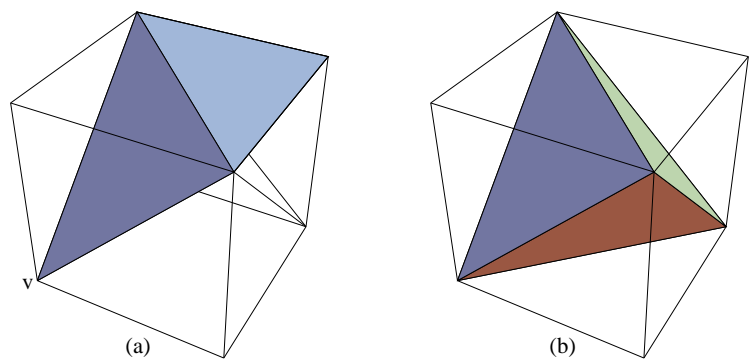}
\caption{(a) One tetrahedron in a triangulation of six tetrahedra starred
from $v$. Here $6$ of the $12$ faces are incident to $v$; $6 = 2n - 10$.
(b) A triangulation using five tetrahedra; the central tetrahedron
is shown.}
\figlab{cubes}
\end{figure}

Their proof follows the structure of Ruppert and Seidel's
similar proof that the same question for nonconvex polyhedra
is intractable~\cite{rs-dttdn-92}.
But the latter authors showed that even deciding whether a polyhedron could
be triangulated is hard, whereas we've seen all convex polyhedra
are easily triangulated.
Here I will sketch just one aspect of the proof in~\cite{blr-cfst-00}.

Both proofs rely on Sch\"{o}nhardt's untriangulable 
polyhedron (\cite[p.~254]{o-agta-87}; \cite[p.~423]{b-t-97}),
shown in Fig.~\figref{Sch}(b).  The three reflex diagonals
(a) block triangulation.  The polyhedron is first transformed by
enlarging the base $B$ (c).
Now if $B$ is glued to a larger ``frame'' polyhedron,
its top face $A$, which BDR call
the ``skylight,'' must be connected through $B$ to a vertex below to
form the tetrahedron that includes $A$.
\begin{figure}[htbp]
\centering
\includegraphics[width=\linewidth]{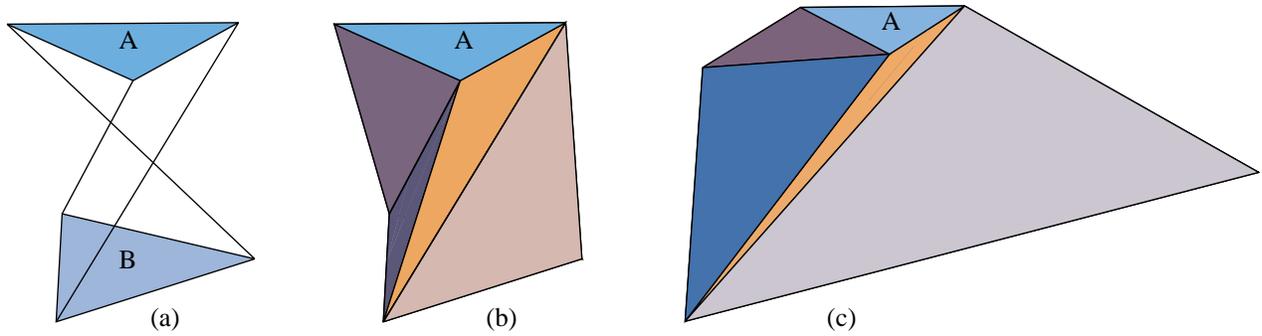}
\caption{Sch\"{o}nhardt polyehdron. (a)
Reflex diagonals;
(b) Polyhedron;
(c) Base $B$ enlarged with respect to ``skylight'' $A$.}
\figlab{Sch}
\end{figure}
It is this ``visibility'' constraint that Ruppert and Seidel
exploited to arrange for their 3-SAT reduction.

BDR convexify the attached Sch\"{o}nhardt polyhedra
through the following strategy.
They prove (in~\cite{bblr-cfst-00})
that a fan-shaped polyhedron
like that shown
in Fig.~\figref{fan} (embedded in a larger polyhedron)
is efficiently triangulated by
employing the internal axis diagonal $ab$, but 
any triangulation that avoids
that diagonal uses many more tetrahedra.
So they string a shallow arc of points exterior to the three reflex diagonals
of each Sch\"{o}nhardt polyhedron, to achieve two goals:
(1) to convexify each; and (2) to heavily penalize any triangulation
that does not employ the reflex diagonals.
This forces the inclusion of the Sch\"{o}nhardt diagonals,
which forces skylight visiblity constraints.
\begin{figure}[htbp]
\begin{minipage}[b]{0.45\linewidth}
\centering
\includegraphics[width=6cm]{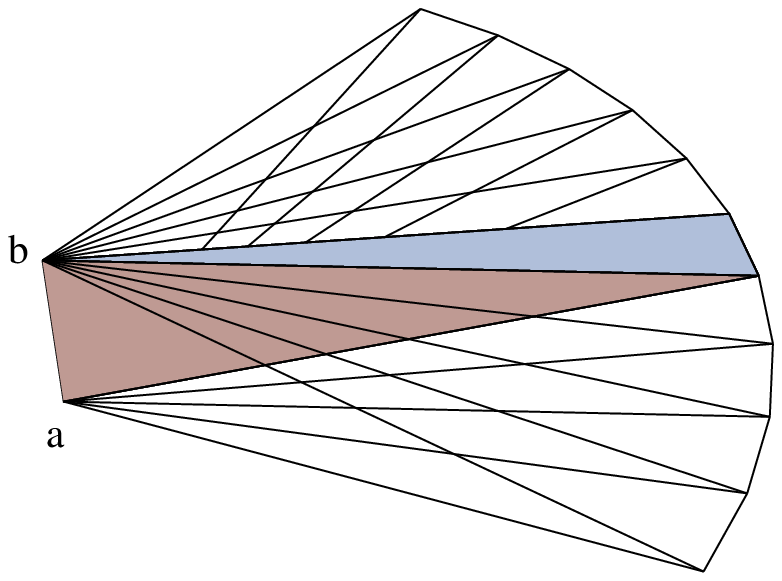}
\caption{A fan is best triangulated using diagonal $ab$; one
tetrahedron is shown.}
\figlab{fan}
\end{minipage}%
\hspace{0.10\linewidth}
\begin{minipage}[b]{0.45\linewidth}
\centering
\includegraphics[height=5cm]{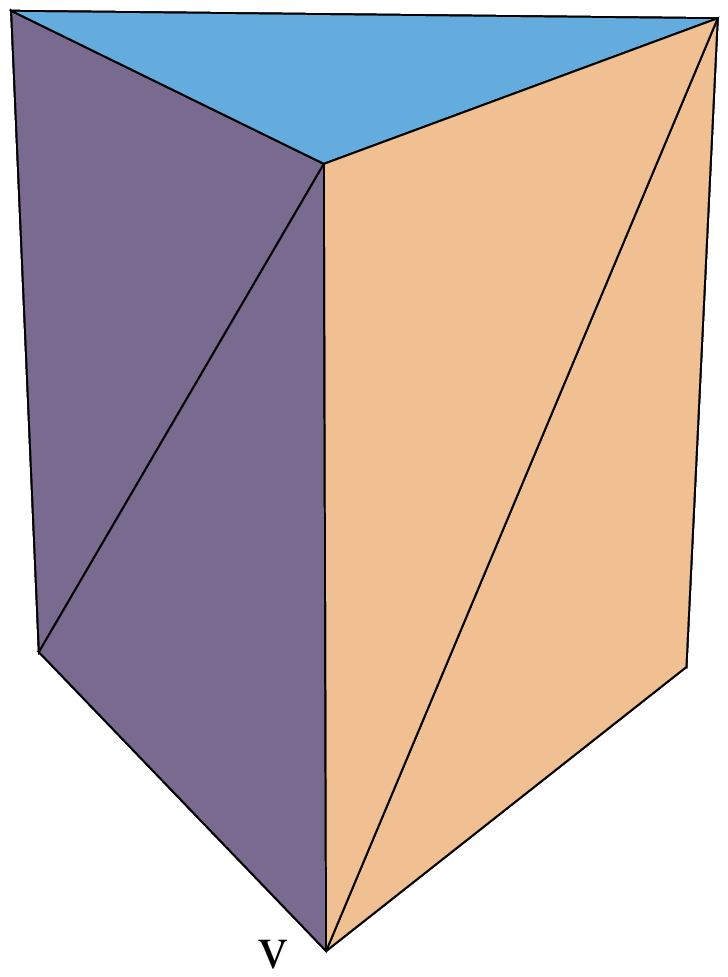}
\caption{Starring from $v$ is incompatible with
surface triangulation.}
\figlab{prism}
\end{minipage}
\end{figure}

The remainder (and majority) of their proof exploits these
constraints to construct variable and clause gadgets
on the frame polyhedron,
carefully arranging lines of sight to result in a
convex polyhedron that can be triangulated with few tetrahedra
iff a particular logical formula is satisfiable.
Beside the intricacy of the logical structure, two
delicate issues are retaining convexity,
and assuring the vertex coordinates remain singly-exponential,
and so polynomially representable.

At least two interesting open questions remain.
The first is determining the complexity of finding a
maximum size (or just a large)
triangulation.  Perhaps surprisingly, large triangulations
of $d$-polytopes
do have application, to algebraic geometry and to integer programming.

The second problem has
practical significance in geometric modeling, in particular,
to meshing.
For a nonsimplicial polyhedron, i.e., one with faces
of more than three sides, the surface may be triangulated
in several different ways.  Typically solid models are
given with a particular surface triangulation.
It is unknown how difficult it is to decide whether there
exists a triangulation of the polyhedron into tetrahedra
that is {\em compatible\/} with a given surface triangulation, in the
sense that each triangle on the surface is a face of a tetrahedron.
For example, no starring of the triangular prism
shown in
Fig.~\figref{prism} is compatible with the displayed (Sch\"{o}nhardt-like)
surface triangulation, for starring from $v$ demands all
faces incident to $v$ be also starred. In fact no triangulation
of this prism is compatible without adding an interior ``Steiner'' point.

\bibliographystyle{alpha}
\bibliography{40}
\end{document}